\documentclass[twocolumn,times,tighten]{aastex631}

\usepackage{color,ulem}
\usepackage{multirow}
\usepackage{amsmath}
\usepackage{mathdots}
\usepackage{bm}
\usepackage{tikz}

\renewcommand{\today}{November 24, 2026}

\shorttitle{Multi-Frequency Inversion of Meridional Flows: 1. Kernels, Method and Validation}
\shortauthors{Thomas Hartlep \& Junwei Zhao}

\begin{document}

\title{Multi-Frequency Inversion of Meridional Flows Using Wave-Simulation-Based Sensitivity Kernels: Kernels, Method and Validation On Synthetic Data}

\correspondingauthor{Thomas Hartlep}
\email{hartlep@baeri.org}

\author[0000-0002-5062-9507]{Thomas Hartlep}
\affiliation{Bay Area Environmental Research Institute, Moffett Field, CA 94035, USA}
\affiliation{NASA Ames Research Center, Moffett Field, CA 94035, USA}

\author[0000-0002-6308-872X]{Junwei Zhao}
\affiliation{W.W. Hansen Experimental Physics Laboratory, Stanford University, Stanford, CA 94305-4085, USA}

\begin{abstract}

Measuring the solar meridional circulation is central to understanding how magnetic flux and angular momentum are redistributed within the Sun.
While time-distance helioseismology has been a primary tool for probing these internal flows, current methods generally rely on broadband travel-time measurements that compress the frequency dependence of the wavefield into single values for each measurement geometry.
This approach overlooks the fact that different temporal frequencies sample the solar interior differently, potentially limiting our ability to resolve complex, depth-dependent flow structures.
In this paper, we present a framework for performing meridional circulation inversions using frequency-dependent travel-time shifts and corresponding sensitivity kernels derived directly from global-scale, 3D wave-propagation simulations.

We analyze sensitivities for horizontal and radial flows, noting that higher temporal frequencies offer increased spatial resolution while their practical application is balanced by the lower signal--to--noise ratio of travel-time measurements at frequencies above the 3 mHz peak of solar oscillations.
Using an iterative regularized least-squares solver, we demonstrate that our method successfully recovers many aspects of complex circulation models while remaining robust against the stochastic noise of realistic simulations.
We consider the impact of different regularization operators on the inversions, and enforcing constraints like mass conservation to improve inversion results.

\end{abstract}


\section{Introduction} \label{sec:introduction}

Time--distance helioseismology is one of the primary tools for probing solar interior flows and wave-speed perturbations by measuring perturbations to acoustic travel times between pairs of surface locations \citep{Duvalletal1993}. 
Typically, cross-covariances are computed from Doppler-velocity (or intensity) signals measured at two points, and acoustic travel times are estimated using either Gabor-wavelet fitting \citep[e.g.,][]{KosovichevDuvall1997, Giles1997} or phase-shift methods \citep[e.g.,][]{GizonBirch2002, ChenZhao2018}. 
These measurements are then interpreted through forward modeling and inversion: travel-time perturbations are related to subsurface perturbations via sensitivity kernels computed in the ray approximation \citep[e.g.,][]{Kosovichev1996, Zhaoetal2001} or in the Born approximation, which accounts for finite-wavelength effects and wave-mode physics \citep{GizonBirch2002, Birchetal2004, Svandaetal2011, Boeningetal2016}. 
More recently, by coupling global-scale wavefield simulations with localized flow perturbations, \citet{HartlepZhao2021} developed a new method for deriving sensitivity kernels for global-scale flows. 
These kernels exhibit an explicit dependence on acoustic frequency, which may enable new inversion strategies for the Sun’s deep meridional circulation that incorporate frequency-dependent travel-time shifts.

Measuring the Sun’s meridional circulation is a central goal of helioseismology because meridional flow plays an important role in transporting magnetic flux and redistributing the Sun’s rotational angular momentum \citep{Miesch2005}. 
Early time--distance studies established poleward near-surface flow and explored its temporal variations using north–south travel-time differences \citep[e.g.,][]{Giles1997, ZhaoKosovichev2004}, motivating increasingly sophisticated inversions based on finite-frequency kernels and improved treatment of observational effects. 
Subsequent work used long time series from SOHO/MDI \citep{Scherreretal1995} and later SDO/HMI \citep{Scherreretal2012, Schouetal2012} to constrain the flow throughout the convection zone, including searches for equatorward return flows and investigations of possible multi-cellular structure \citep[e.g.,][]{Zhaoetal2013, Kholikovetal2014, Jackiewiczetal2015, RajaguruAntia2015, ChenZhao2017, LinChou2018, Liangetal2018}.

A persistent challenge has been isolating the relatively small meridional-flow signal from systematic effects, including center-to-limb effect \citep{Zhaoetal2012, Baldner2012, ChenZhao2018}, magnetic contamination \citep{LiangChou2015, ChenZhao2017}, foreshortening \citep{Zhaoetal2016}, and measurement differences arising from different instruments \citep{Gizonetal2020, Braunetal2021}. 
These effects can rival or exceed the expected travel-time signatures of deep flows, motivating improved calibration strategies, empirical corrections, and forward modeling that better capture the measurement process. Although the present paper does not directly address these observational or data-analysis systematics, the framework developed here provides another inversion approach that may help mitigate one or more of the issues noted above.

A notable commonality across essentially all prior time–distance inversions for meridional circulation is that the inversion input consists of broadband travel-time shifts: for each measurement geometry (e.g., distance, annulus, quadrant), a single travel-time perturbation (or a small set of travel-time definitions) is obtained by fitting the cross-covariance over a wide frequency band.
The corresponding kernels are computed for that same measurement definition, yielding an integral relationship between a scalar travel-time shift and subsurface flow.
While powerful, this approach compresses the frequency dependence of the wavefield into a single value for each geometry.
As a result, it does not directly exploit the fact that different temporal frequencies, and thus different mode content, sample the solar interior differently, nor does it provide full information on the solar interior dynamics.

In an earlier paper, \citet{HartlepZhao2021} derived frequency-dependent sensitivity kernels for global-scale flows. 
Here we present a framework for performing meridional-circulation inversions using frequency-dependent travel-time measurements together with improved kernels, and we validate the analysis procedure using a two-stage approach: first testing against forward models and then against synthetic measurements from wavefield simulations.
This controlled validation ensures that our multi-frequency framework is robust before being applied to real solar data, which remains confounded by significant systematic effects like center-to-limb variations that have not been fully understood yet.

The remainder of this paper is organized as follows.
Section~\ref{sec:sensitivity_kernels} describes the sensitivity kernels. 
Section~\ref{sec:models} introduces the meridional-circulation models and wavefield simulations. 
Section~\ref{sec:inversion_procedure} presents the frequency-dependent measurements and inversion procedures. 
Section~\ref{sec:synthetic} discusses inversions using synthetic measurements, and Section~\ref{sec:summary} provides a summary and conclusions.

\newpage
\section{Improved Sensitivity Kernels} \label{sec:sensitivity_kernels}

\begin{figure*}[p]
   \centering
   \includegraphics[width=0.97\linewidth]{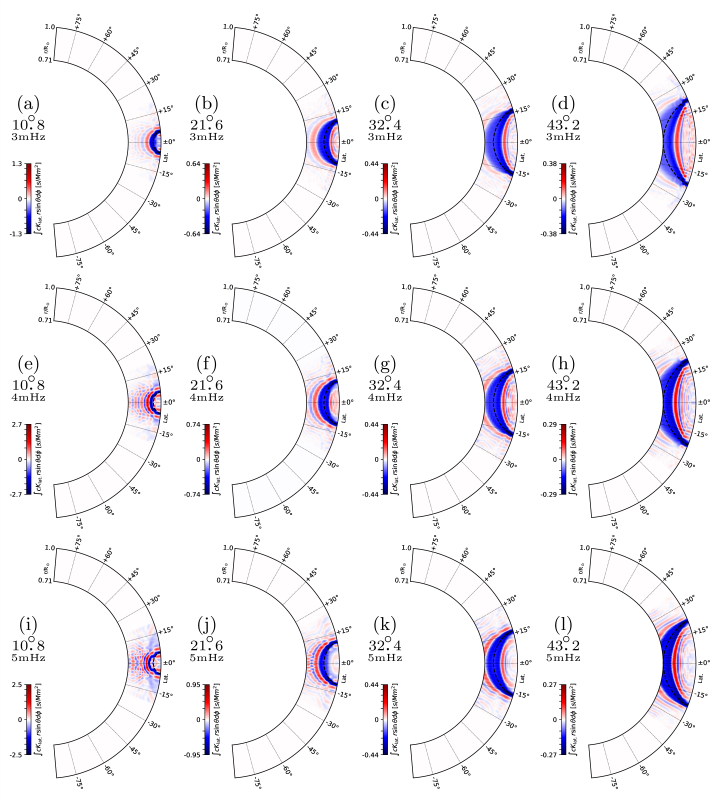}
   \caption{Azimuthally-integrated sensitivity kernels of North-South wave travel-time differences to latitudinal flows.
   Shown are results for different central frequencies $\nu=3$, 4 and 5~mHz (from top to bottom) and  travel distances  $d=10\fdg8$,  $21\fdg6$, $32\fdg4$ and $43\fdg2$ (from left to right), respectively.
   Dashed curves in each panel indicate the wave path in the ray approximation.
   The shown kernels are centered at the equator, and correspond to travel-time differences measured between symmetric latitudes at $\pm d/2$.
   The kernel $K_{\rm lat}$ as shown has been multiplied with the local sound speed $c$ in order to make the kernel in the deeper interior more easily visualized since its amplitude reduces strongly with depth.
       \label{fig:horizontal_kernels}}
\end{figure*}
\begin{figure*}[p]
   \centering
   \includegraphics[width=0.97\linewidth]{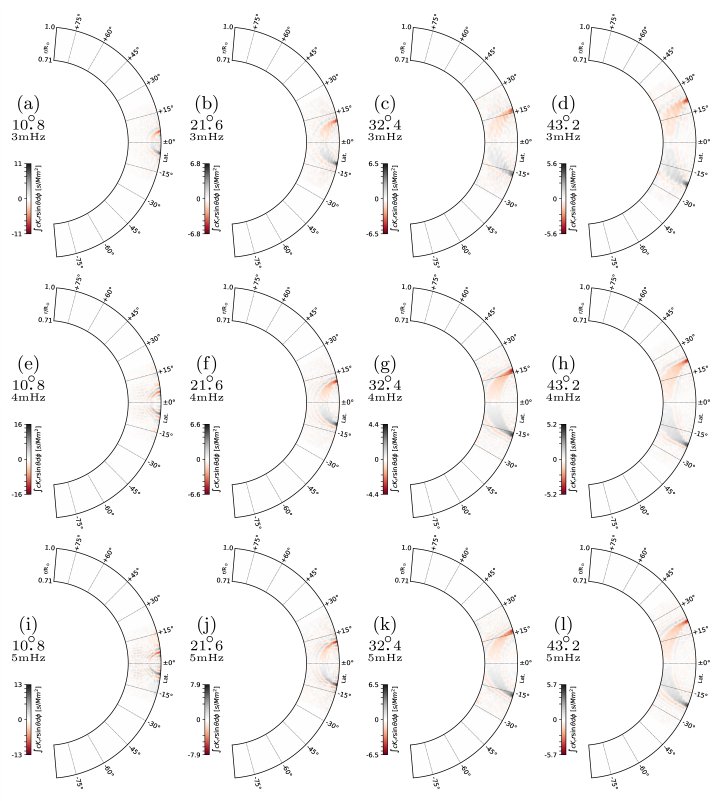}
   \caption{Azimuthally-integrated sensitivity kernels for radial flows similar to Figure~\ref{fig:horizontal_kernels}.
   The panels show kernels centered at the equator for frequency bins centered at $\nu=3$, 4 and 5~mHz (from top to bottom) and for four travel distances,  $d=10\fdg8$,  $21\fdg6$, $32\fdg4$ and $43\fdg2$ (from left to right).
   Again, the shown kernels are centered at the equator, corresponding to travel-time differences measured between symmetric latitudes $\pm d/2$, and are multiplied by the sound speed $c$ to enhance visibility in the deep interior.
       \label{fig:radial_kernels}}
\end{figure*}

Helioseismic sensitivity kernels relate travel-time perturbations measured at the solar surface to subsurface flows and wave-speed perturbations.
Traditional kernel calculations have relied on ray or Born approximations.
In \citet{HartlepZhao2021}, we introduced an alternative approach based on numerical wave-propagation simulations, from which frequency-dependent sensitivity kernels can be inferred directly.
Here we present refined versions of those kernels and extend the method to radial flows.

In this paper, we reuse the simulations and measurements from \cite{HartlepZhao2021} but recompute the kernels at substantially increased radial resolution and accuracy.
Previously, in order to save computational cost and reduce memory requirements, we used a radial resolution that was approximately the same as the horizontal resolution.
This resolution is lower than what is used in the wave-field simulations, which require much higher resolutions radially in order to resolve the strong radial gradients in the Sun.
The new kernel calculations presented here use the full radial resolution from the wave propagation simulations.
This greatly increases the size of the linear systems to be solved.
For example, for the horizontal kernels for a travel distance of $21\fdg6$, the linear system has a size of
$15,599,528 \times 10,301,424$ compared to $2,390,444 \times 2,243,052$ for the resolution used in \cite{HartlepZhao2021}.
Storing and solving such large matrices is only possible by exploiting, and in fact deliberately increasing the sparseness of the problem by neglecting matrix elements associated with flow amplitudes smaller than some threshold.
\cite{HartlepZhao2021} used a threshold fraction of $10^{-3}$ of the maximum perturbation velocity, while the present calculations adopt a smaller threshold of $10^{-4}$, retaining more non-zero matrix elements.
We use sparse matrix storage and the sparse iterative least-squares solver LSQR \citep{PaigeSaunders1982a} available from the Python library SciPy \citep{2020SciPy-NMeth}.

In the new calculations, we also revise the form of the regularization.
In our previous work \citep{HartlepZhao2021}, the regularization penalized the overall amplitude of the kernel by minimizing the volume integral of its squared magnitude (with appropriate weighting for the spatial discretization).
This approach suppresses large values everywhere but enforces smoothness only indirectly.
In the present formulation, we instead penalize spatial curvature by applying a Laplacian-based regularization operator, discretized using fourth-order finite differences.
This directly constrains second-order spatial variations of the kernel, leading to smoother solutions while avoiding unnecessary damping of the amplitude.

The strength of the regularization is controlled by a scalar parameter, $\gamma$.
As $\gamma$ is decreased, the residuals initially decline, reach a distinct minimum, and then increase again as the solution begins to compensate for errors arising from linearization and discretization.
We therefore adopt the value at this minimum as the optimal regularization parameter.
We note that this behavior differs from the more familiar L-curve tradeoff encountered in the flow inversions using fixed kernels (Section~\ref{sec:inversion_procedure}), where the residual decreases monotonically with decreasing regularization.
This difference reflects the dominant error sources in the two problems: realization noise in the travel-time measurements, and linearization and discretization errors in the kernel calculations.

In addition to the improved kernels for horizontal flows, we have also repeated the procedure for radial flows using the same simulation setup, perturbation strengths, and measurement and inversion procedures.
This involved simulating wave propagation through 784 individual 3-D Gaussian-shaped radial flow perturbations and measuring the resulting change in wave travel time between wave source and all points on the solar surface.
We then derived the sensitivity kernels for radial flows by again solving the coupled set of equations in a least-squares sense, as in \cite{HartlepZhao2021}.
Here and in \cite{HartlepZhao2021}, we compute these three-dimensional kernels on a locally defined discrete spherical grid in radial, longitudinal and transverse directions.

For the purpose of inferring meridional flows, we do not use these full three-dimensional kernels directly.
Instead, we employ azimuthally integrated kernels and corresponding azimuthally averaged travel-time measurements.
Figures~\ref{fig:horizontal_kernels} and~\ref{fig:radial_kernels} show some of these integrated kernels for select travel distances.
They have been computed by integrating the discrete kernels over the transverse direction, defined as the horizontal direction perpendicular to the wave-propagation path on the spherical surface.
Although this is not exactly equivalent to integrating over solar latitude, the difference arises because the transverse direction follows a locally propagation-aligned coordinate rather than the global spherical coordinate system; however, the discrepancy is small since the kernels are strongly localized near zero transverse offset.
The figures show integrated kernels centered at the equator, that is, they correspond to travel-time measurements between latitudes $+d/2$ and $-d/2$, where $d$ is the travel distance.
When performing inversions, we shift these kernels to the central latitudes of the different travel-time measurements.
As in \cite{HartlepZhao2021}, separate kernels have been computed for different frequency bins centered at 3, 4 and 5~mHz with a bin width of 0.33~mHz.
The sign of the kernels has been chosen such that the kernels correspond to the travel-time shifts experienced by northward traveling waves given a northward flow (for the latitudinal kernel) or upward flow (for the radial kernel).
The figures show the kernels multiplied with the local sound speed $c$ in order to better visualize the low amplitude contributions in the deep interior.

From these plots, some properties of the sensitivity kernels are apparent.
First, the kernels for latitudinal flows are symmetric with respect to the wave travel path, whereas radial kernels are antisymmetric.
As a result, spatially uniform flows in latitudinal direction produce nonzero travel-time shifts, while radial flows need to vary spatially to yield a net effect.
Second, the radial kernels exhibit larger amplitudes than the horizontal kernels by almost an order of magnitude, which may appear counterintuitive given that radial flows generally produce smaller travel-time shifts.
However, travel-time shifts are a consequence of the product of flow speed and the kernel values, and radial flows tend to be much weaker in the Sun than horizontal flows because of strong radial stratification and mass conservation.
The antisymmetric nature of the radial kernels also means that travel-time shifts for large-scale radial flows tend to cancel out at least partially.
Third, both types of kernels show characteristic ``shell-like'' regions of alternating sign around a primary region of sensitivity which approximately aligns with the ray path as indicated by dashed curves in Figure~\ref{fig:horizontal_kernels}.
The full 3-D kernels for horizontal flows (not shown here) do exhibit a significant reduction in sensitivity right along the ray path, as did the kernels in our previous paper \citep{HartlepZhao2021}, but this is not apparent in the azimuthally-integrated kernels shown here.
Radial-flow kernels do not appear to show any significant reduction along the ray path.
As for other similar kernels, kernels for higher frequency waves show more closely spaced shells than kernels for lower frequency waves, and the central region of sensitivity tends to be more spatially confined at higher frequencies.

Formally, we have computed time-distance helioseismic sensitivity kernels for 60 different travel distances from $2\fdg16$ to $44\fdg64$ (heliographic distance between points on surface) with a spacing of $0\fdg72$, and for three distinct frequency bins of 3, 4, and 5~mHz.
However, we exclude the shortest travel distances from the inversions, as the kernels are likely inadequately resolved in the underlying wave-propagation simulations, and also appear inconsistent with travel-time measurements for two test flows.
This limitation arises because short-distance waves turn at shallow depths, where the sound speed is relatively low.
For a given wave frequency, this implies correspondingly short horizontal wavelengths and thus high spherical harmonic degrees needed to resolve them.
At a travel distance of $10.8\degr$, given a wave frequency of 3~mHz, the required spherical harmonic degree is $\ell \sim 130$, which is only marginally resolved given the dealiased resolution of the simulations ($\ell_{\max} \sim 170$).
At shorter distances, the resolution requirements become more severe: $\ell \gtrsim 160$ at $8\degr$ and $\ell \gtrsim 250$ at $6\degr$, exceeding the effective resolution.\footnote{These estimates are based on the lower turning depth and local sound speed along the ray path; e.g., at $10.8\degr$ the turning point lies at $\sim 47~\mathrm{Mm}$ where $c_s \approx 93~\mathrm{km~s^{-1}}$, corresponding to a horizontal wavelength of $\sim 31~\mathrm{Mm}$.}
For higher wave frequencies, these resolution limits become even more stringent.
In contrast, waves associated with large travel distances propagate through the near-surface, low-sound-speed layers primarily in the radial direction and only acquire significant horizontal structure near their deeper turning points, where the sound speed is higher and the corresponding horizontal wavelengths are sufficiently large to be well resolved.
The radial resolution, set by the need to resolve the strong stratification of the solar interior, remains adequate for all waves considered.
We therefore attribute the inconsistency of the shortest-distance kernels to insufficient horizontal resolution in the forward simulations.
The inversions below therefore only use kernels for distances $\ge 10\degr$ (for 3, and 4~mHz), and $\ge 16\degr$ (for 5~mHz).

\section{Flow Models and Simulations} \label{sec:models}

In this study, we have used two different flow models and synthetic data from corresponding helioseismic wave propagation simulations for developing and validating our meridional flow inversion code.

We prescribe these models as background flows in our three-dimensional linearized wave propagation code \citep{HartlepMansour2005, Hartlepetal2013} in order to compute the propagation of helioseismic waves through it.
Unlike the simulation for deriving sensitivity kernels, where we simulate the propagation of single waves started from initial conditions, we here simulate the temporal evolution of the stochastic helioseismic wavefield.
Waves are excited by stochastically driving density perturbations just below the photosphere ($\sim150$km below), similar to the Sun, and with a temporal and spatial spectrum such that the resulting wave power spectrum is solar-like.
From these simulations, we extract synthetic helioseismic observations, specifically the radial oscillation velocity at 300~km above the photosphere.
We use these data to measure travel-time shifts for waves travelling along the latitudinal direction.
This follows the frequency-dependent procedure described by \citet{ChenZhao2018}, with one key difference. 
In their analysis, the frequency-dependent shifts are defined as differences between travel times measured for oppositely propagating wave directions.
In our synthetic-data analysis, however, we use corresponding sets of simulations without prescribed flows as reference models, and measure travel-time shifts relative to the reference.
This helps reduce the effect of stochastic noise in the simulations on the measurements.
These measurements are multiplied by two such that they correspond to the north-south travel-time differences measured observationally.
Finally, the resulting travel-time differences are averaged azimuthally.

\subsection{Single-Cell Model}

\begin{figure}[t]
  \includegraphics[width=\linewidth]{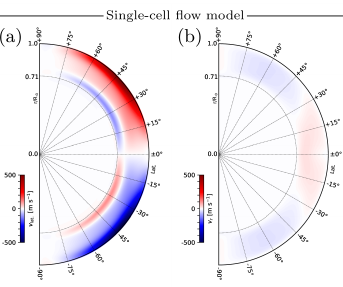}
   \caption{Prescribed flows for the single-cell model used in this study, with panels (a) and (b) showing the latitudinal and radial components, respectively.
       \label{fig:single_cell_flow_model}}
\end{figure}

\begin{figure*}[p]
  \includegraphics[width=\linewidth]{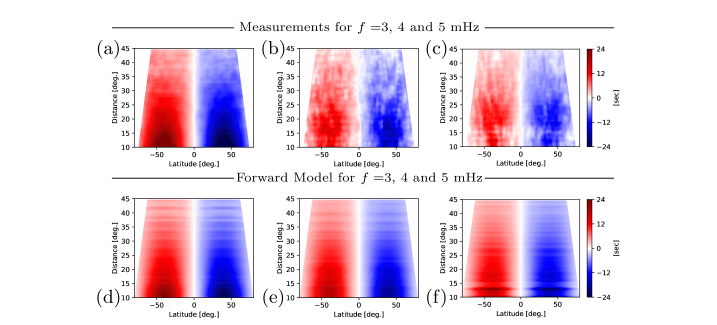}
   \caption{Travel-time shifts for the single-cell model.
   Panels (a--c) show travel-time measurements from the numerical simulation with a prescribed single-cell flow, for the three frequency bins considered.
   Panels (d--f) show the results of corresponding forward calculations using the simulation-based sensitivity kernels.
       \label{fig:single_cell_measurements}}
\end{figure*}

\begin{figure*}[p]
   \centering
   \includegraphics[width=\linewidth]{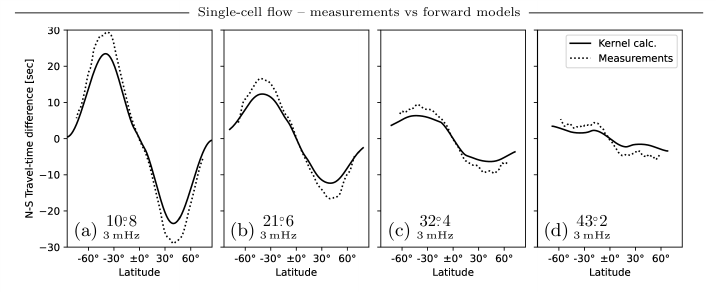}
   \caption{Comparing measurements and forward models for the single-cell flow.
       Shown are the north-south travel-time shifts at four different travel distances as indicated on the plot, all for the 3\,mHz frequency bin.
       Dotted curves denote helioseismic measurements on the simulated wavefield, while solid curves show forward calculations using the simulation-based sensitivity kernels.
       \label{fig:single_cell_measurements_vs_forward}}
\end{figure*}

The first model we consider was taken from the mean field differential rotation and meridional circulation model of \citet{Rempel2006}, of which we only use the meridional flow here.
This flow is characterized by a single meridional circulation cell in each hemisphere, and we therefore refer to it as the ``single-cell'' flow model.
It has both latitudinal and radial components which obeys mass conservation.
\cite{Hartlepetal2013} used this model to produce a 4096~min long dataset with 1~min cadence, which we reuse here.
A corresponding simulation without meridional flow was used as a reference for measuring phase and corresponding travel-time shifts.

For solar observations, usually many months to years of data are used when inferring meridional flows due to the statistical nature of solar oscillations and the small effect meridional flows have on travel times.
Such long-duration simulations are still not possible, however.
In order to make shorter numerical simulations in a sense equivalent to longer-duration ones, \cite{Hartlepetal2013} significantly increased the flow amplitude of the model to a peak of 500~ms$^{-1}$ (approximately a factor of 36 increase).
This is still significantly smaller than the sound speed anywhere in the solar interior.
It therefore does not change the physics of the problem but merely provides a large increase in the signal--to--noise (S/N) ratio of the resulting travel-time shifts.
Since the (uncorrelated) realization noise reduces with the square root of the length of a time series, such simulations should have an S/N that is equivalent to a simulation of the original model but run for a period $\approx36^2$ times longer.

Figure ~\ref{fig:single_cell_flow_model} shows the radial and latitudinal components of the model as run in the simulation.
New measurements of the travel-time differences between northward- and southward-traveling waves from this dataset are shown in Figure~\ref{fig:single_cell_measurements}a--c, for the three frequency bins considered.
While the 3~mHz measurements are fairly smooth, more significant stochastic noise is visible in the measurements for higher frequencies.
We also compute the forward model, that is, the expected travel-time shifts given the flow model and our sensitivity kernels, shown in Figure \ref{fig:single_cell_measurements}d--f.

Measurements and forward models match well in their latitudinal and distance dependence, showing the largest amplitudes at mid latitudes in each hemisphere, with travel-time shifts decreasing with increasing travel distance.
The 5~mHz forward model shows spurious values for travel distances $\lessapprox16\degr$, which we attribute to the resolution limit of the kernel simulations (see discussion in section~\ref{sec:sensitivity_kernels}). 
Consequently, we will exclude these distances in the inversions using this frequency.

Figure~\ref{fig:single_cell_measurements_vs_forward} shows more detailed comparisons for selected travel distances, and reveals a sizable offset between forward model and measurements.
While the origin of this discrepancy is not truly understood, it is notable that the match between measurements and forward models is much closer for the double-cell model (see Figure~\ref{fig:double_cell_measurements_vs_forward}) for which we have three separate simulations with different noise realizations while the single-cell flow measurements use only a single simulation.

\subsection{Double-Cell Model}

\begin{figure*}[p]
   \centering
   \includegraphics[width=0.97\linewidth]{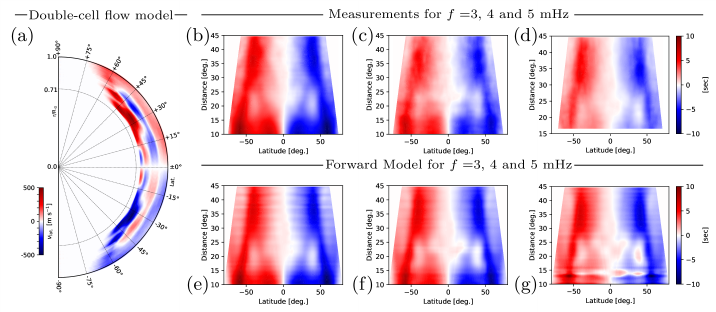}
   \caption{Similar to Figure~\ref{fig:single_cell_flow_model}, panel (a) shows the prescribed latitudinal flow for the double-cell model used here. The radial flow is zero for this model.
   Panels (b--d) show travel-time measurements using the frequency-dependent measurement technique for 3, 4 and 5~mHz frequency bins.
   Panels (e--g) show the results from corresponding forward calculations.
       \label{fig:double_cell_flow_model}}
\end{figure*}

The second flow model is derived from the meridional flow inversion of \cite{Zhaoetal2013}
which is characterized by two circulation cells in each hemisphere.
In the following, we refer to this model as the ``double-cell'' flow model.
Compared to the results published in \cite{Zhaoetal2013}, we have however made some modifications.
Again, we increased the maximum flow speed to 500~ms$^{-1}$ to increase the S/N ratio.
We also modified the speed of the bottom circulation cell by a factor of two.
This model consists of only the latitudinal flow component, and does not strictly obey mass conservation.

Figure ~\ref{fig:double_cell_flow_model}a shows the latitudinal component of the model as run in the simulation.
The radial flow is zero in this case.
With this model, we ran three separate simulations with different statistical realizations of the forcing term, each producing 48~hour-long datasets.
The first 4~hours of each simulation were discarded as they represent transient behavior.

Using the frequency-dependent measurement procedure, phase-shifts and corresponding travel-time shifts were derived from these synthetic data.
Figure~\ref{fig:double_cell_flow_model}b--d show the travel-time shifts, for the three frequency bins (3, 4 and 5~mHz), averaged over the three separate datasets, and averaged azimuthally.
Combining these independent simulations increases the S/N of measured travel times, and allows for uncertainty estimation by computing the standard deviation between these sets.
However, because three datasets only give a poorly constrained variance for each individual measurement, we pool the variances of all measurements within $\pm 7.5\degr$ in latitude and $\pm 1.5\degr$ in travel distance, after
removing the per-measurement mean across datasets. 

Corresponding forward models are shown in Figure~\ref{fig:double_cell_flow_model}e--g.
Measurements and forward models show very similar patterns in distance and latitude, in particular for the frequency with strongest power (3~mHz), while discrepancies start to increase for the higher frequency bins.
In the 5~mHz forward model there are again some spurious features at $\sim 13\degr$ travel distance, a sign that these kernels may be less reliable and possibly marginally resolved.

\begin{figure*}[p]
   \centering
   \includegraphics[width=\linewidth]{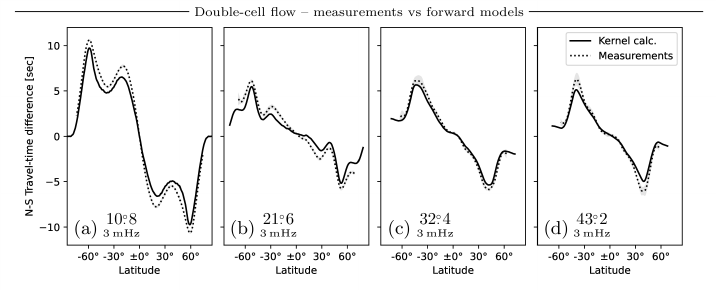}
   \caption{Comparing measurements and forward models for the double-cell flow.
       Shown are the north-south travel-time shifts at four different travel distances as indicated on the plot, all for the 3\,mHz frequency bin, with 
       dotted curves denoting helioseismic measurement on the simulated wavefield, and solid curves denoting forward calculations using the simulation-based sensitivity kernels.
       For the measurements, pooled statitical uncertainties are shown by shading around the curves, corresponding to one standard deviation.
       \label{fig:double_cell_measurements_vs_forward}}
\end{figure*}

Figure~\ref{fig:double_cell_measurements_vs_forward} shows travel-time shifts for a select number of travel distances, comparing in more detail the measurements and forward model.
They show very good correspondence between measurement and forward model, much closer than the single-cell comparison.
Stochastic noise in these measurement data is small at 3~mHz but becomes noticeably larger for higher frequencies (not shown).

\section{Inversion Procedure} \label{sec:inversion_procedure}

\subsection{Basic Problem} \label{sec:inversion_procedure:basicproblem}

Similar to the way we infer sensitivity kernels from known flow perturbations and the travel-time shifts they cause, we can determine the flows in the solar interior given kernels and observed travel-time perturbations.
Again, this is solved on a discrete grid.
Specifically, we use 243 latitudinal grid points spanning approximately $[-85\degr, +85\degr]$ with an equidistant spacing of $\sim0\fdg7$, and 174 radial grid points spanning $[0.71, 1]\,r/R_\sun$ with a stretched spacing that is approximately proportional to the depth-varying sound speed in the Sun.
In discrete form, the convolution equation that relates flow velocity to travel-time shifts becomes a sum:
\begin{equation}
   \delta\tau_i = \sum_j \vec{K}_{i,j} \vec{u}_j S_j,
   \label{eqn:discretesum}
\end{equation}
where $\delta\tau_i$ are the azimuthally-averaged N-S travel-time differences, $\vec{K}_{i,j}$ are the corresponding azimuthally-integrated kernels, and $\vec{u}_j$ is the azimuthally-averaged flow.
$S_j$ is the integration element such that this sum corresponds to the integral over radius $r$ and co-latitude $\theta$, that is $S_j=r\,dr\,d\theta$.
$dr$ and $d\theta$ are computed using 4-th order finite differences.
Index $i$ in the equation labels the different measurements (distances and central latitudes), while $j$ is a single index over all the grid points of our domain.
Note that $\vec{K}$ and $\vec{u}$ may have more than one component, e.g. both latitudinal and radial flows.
The coupled set of Eqn.~\ref{eqn:discretesum} for all $i$ is a matrix equation
\begin{equation}
Ax=b
\label{eqn:matrix}
\end{equation}
where $A$ contains the kernel and integration elements, and $b$ contains the measurements.

\begin{figure*}[t]
   \centering
   \includegraphics[width=\linewidth]{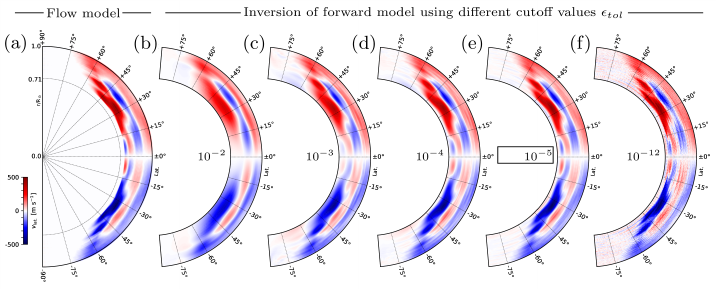}
   \caption{Inversion of forward calculations for the double-cell flow model performed without an explicit regularization term. Panels (b)--(f) show the inverted latitudinal flow $v_{\rm lat}$ solved using different values of the cut-off parameter which determines early stopping of the iterative solver. For comparison, panel (a) shows the actual flow model. The same color scale is used for all panels.}
   \label{fig:differentcutoffparameters}
\end{figure*}

In our setup, Eqn.~\ref{eqn:matrix} is underdetermined, there are more unknowns (the flow field) than knowns (the measurements), and we cannot solve the equation directly.
However, we can compute least-squares solutions to it, that is, we minimize $||Ax-b||$ where $||\cdot||$ is the Euclidean norm.
As a first test we may use, instead of actual measurements, the expected travel-time differences (the forward model) for a given flow model.
Specifically, unless explicitly stated, we use the forward models computed using the 3~mHz kernels.

In general, solving this system requires some form of regularization in order to compute physically meaningful, reasonably smooth solutions.
However, since there is no noise in the forward model, it is exact within machine accuracy, very little regularization is needed here.
Using LSQR, an iterative least-squares solver, we can in fact solve Eqn.~\ref{eqn:matrix} without needing an explicit regularization term.
Using different cut-off parameters $\epsilon_{tol}$, we stop iterations once the residual $||Ax-b|| < \epsilon_{tol} (||A| ||x|| + ||b||)$.
Figure~\ref{fig:differentcutoffparameters} shows the result of inversions for our double-cell flow model, for a range of cut-off parameters.
Large values result in early stopping and smoothed inversions, while the flow is progressively more accurately recovered as the cut-off parameter is reduced.
Noise becomes visible for extremely small values of $\epsilon_{tol}$.
A good compromise is $\epsilon_{tol}=10^{-5}$ which results in an excellent solution.

\subsection{Explicit Regularization} \label{sec:inversion_procedure:explicitregularization}

\begin{figure*}[t]
   \centering
   \includegraphics[width=\linewidth]{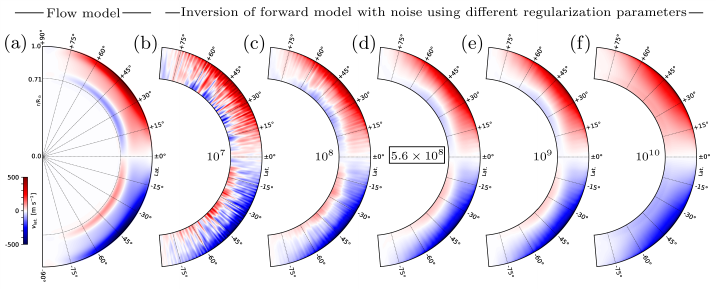}
   \caption{Comparison of the effect of different regularization parameters on the inversion of the single-cell meridional flow model using the depth-scaled Laplacian $s_{10}\Delta$ as the regularization operator.
   Panel (a) shows for reference the latitudinal flow speed of the flow model itself while panels (b)--(f) show results of inversions of the forward model with 5\% Gaussian noise added.
   For this case, only the latitudinal flow is used in the forward model and inversion.
   All panels use the same color scale as shown on the far left.
   Following an L-curve argument (see figure~\ref{fig:single_cell__inversion_of_forward_model_with_noise__L_curve}), the optimal regularization parameter for this case is $\gamma_{\text{L}}=5.6\times10^8$.
   \label{fig:single_cell__inversion_of_forward_model_with_noise__different_regularization_parameters}}
\end{figure*}

\begin{figure}[t]
   \centering
   \vspace*{5mm}
   \includegraphics[width=0.95\linewidth]{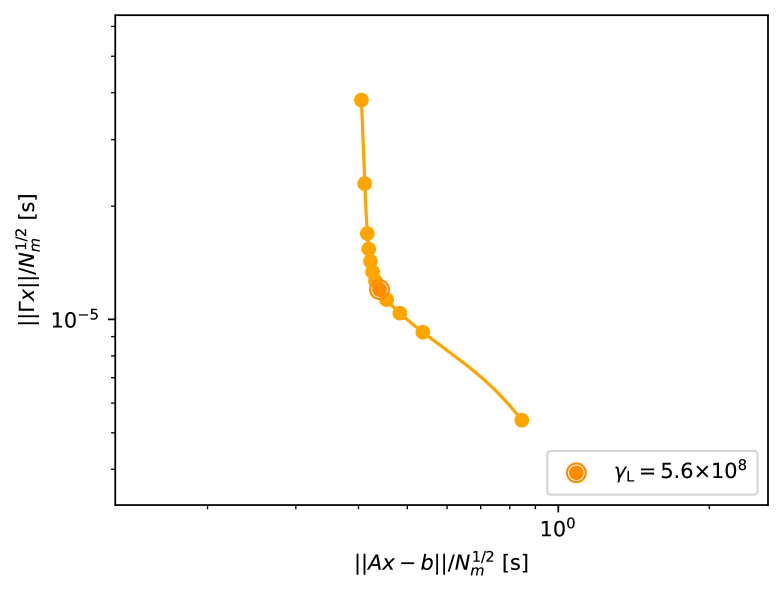}
   \caption{L-curve for the inversions of the single-cell forward model with Gaussian noise from Figure~\ref{fig:single_cell__inversion_of_forward_model_with_noise__different_regularization_parameters}, illustrating the trade off between inversion residual, $||Ax-b||$, and regularization as measured by $||\Gamma x||$. 
   The residuals shown are divided by the number of measurements, $N_m$, providing a measure of the average misfit per measurement.
   The curve is traversed from right to left as the regularization parameters is decreased.
   \label{fig:single_cell__inversion_of_forward_model_with_noise__L_curve}}
\end{figure}

\begin{figure*}[t]
   \centering
   \includegraphics[width=\linewidth]{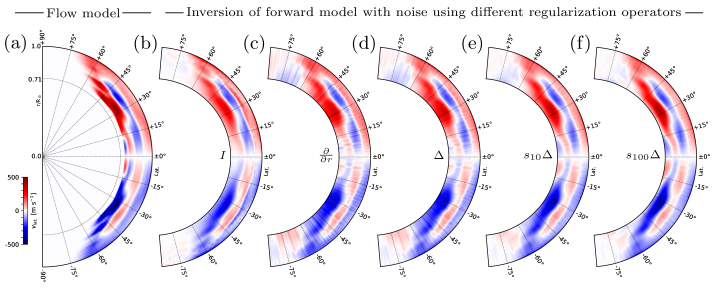}
   \caption{Comparison of the effect of different regularization operators on the inversion of the double-cell meridional flow model.
   For reference, panel (a) shows the latitudinal flow speed of the model while panels (b)--(f) show results of noise-weighted inversions of a forward model with 5\% Gaussian noise.
   For each case, the final regularization parameter was selected by the discrepancy principle with the following values: $\gamma_{I}\approx1.7\times10^5$, $\gamma_{\frac{\partial}{\partial r}}\approx5.6\times10^7$, $\gamma_{\Delta}\approx5.1\times10^7$,  $\gamma_{s_{10}\Delta}\approx1.6\times10^7$,  $\gamma_{s_{100}\Delta}\approx5.4\times10^5$.
   All panels use the same color scale shown in the far left panel.
   \label{fig:double_cell__inversion_of_forward_model_with_noise__different_regularization_operators}}
\end{figure*}

Instead of the regularization offered by early stopping, inversion of measurement data requires an explicit regularization term to obtain stable and physically meaningful solutions.
Such regularization compensates for both statistical noise in the measurements (errors in $b$)
and systematic differences between the approximate sensitivity kernels and the true physical response of the system (errors in $A$).
We therefore solve the modified least-squares problem
\begin{equation}
  \min_{x} \left( ||Ax-b||^2 + \gamma||\Gamma x||^2 \right),
\end{equation}
where $\gamma$ is a scalar regularization parameter, and $\Gamma$ is a discretized regularization operator which also contains discrete integration elements $S_i$ such that $||\Gamma x||^2$ corresponds to the azimuthal integral of the squared regularization term.
This expression can also be rewritten in augmented matrix form as
\begin{equation}
\left[ {\begin{array}{c}
   A \\
   \sqrt{\gamma} \Gamma \\
 \end{array} } \right]
 x =
 \left[ {\begin{array}{c}
     b \\
     0 \\
   \end{array} } \right].
   \label{eqn:augmented_system}
\end{equation}

We first illustrate the effect of $\gamma$ using synthetic measurements, consisting of forward-modeled travel-time perturbations generated from the single-cell meridional-flow model using the numerical kernels.
To assess the robustness of the inversion procedure under stochastic perturbations, we add Gaussian noise at a 5\% level.
Specifically, each synthetic measurement is perturbed by an independent normally distributed random value with a standard deviation equal to 5\% of the corresponding measurement amplitude.
Because the perturbations are uncorrelated between measurement locations and travel distances, this represents a relatively stringent test of the inversion stability.
Figure~\ref{fig:single_cell__inversion_of_forward_model_with_noise__different_regularization_parameters} shows the resulting inversions for different values of $\gamma$, for one particular choice of regularization operator.
The inversion results are very noisy when $\gamma$ is small, and become increasingly smoother as $\gamma$ is increased.

The solution thus depends strongly on how the balance between data misfit and model smoothness is set, and the question arises how to select $\gamma$ appropriately.
We consider two criteria: the L-curve, which does not require knowledge of the noise level, and the discrepancy principle, which does.

The L-curve \citep{Hansen1992, HansenOLeary1993} plots the regularization seminorm, $||\Gamma x||$, against the residual norm, $||Ax-b||$, on logarithmic axes.
As $\gamma$ is decreased, the residual decreases monotonically while the seminorm increases, slowly at first and then sharply once noise starts to be fitted.
The resulting curve bends at a corner, which represents in some sense an optimal compromise between stability and resolution.
We locate the corner by computing inversions for a set of $\gamma$ values spaced logarithmically over a wide range and finding the point of maximum curvature.
Figure~\ref{fig:single_cell__inversion_of_forward_model_with_noise__L_curve} shows the L-curve for the single-cell test case.
Because the true solution is smooth and the noise is strong, the seminorm is not constant on the branch of large $\gamma$ but varies noticeably there, yet the corner remains clearly identifiable.
In general, however, the L-curve criterion is heuristic and the corner can be poorly defined, in which case it may over- or under-regularize.
Its location can also depend on the choice of regularization operator and on correlations in the noise.

When the noise level is known, a more direct criterion for selecting $\gamma$ is the discrepancy principle \citep{Hansen1992, HansenOLeary1993}.
It states that the solution should not fit the data more closely than the noise level allows.
Suppose the data were generated by a known model, $b=Ax_{\rm true} + \epsilon$, where $\epsilon$ is uncorrelated noise of variance $\sigma^2$ in each of the $N_m$ measurements, as in our synthetic tests.
The residual of the true model is then exactly the noise, and its squared norm is on average $N_m\sigma^2$.
Although $x_{\rm true}$ is unknown in practice, this sets the expected residual of an acceptable solution: substantially smaller residuals indicate that noise is being fitted (underregularization), while substantially larger ones indicate that signal is being suppressed (overregularization).
The discrepancy principle therefore selects the $\gamma$ for which the regularized solution, $x_\gamma$, satisfies
\begin{equation}
   ||Ax_\gamma-b||^2 \approx N_m \sigma^2 .
\end{equation}
When measurements have different uncertainties $\sigma_i$, as is the case for our noise model, we can divide each equation by its $\sigma_i$, i.e. $\tilde{A}_{ij}=A_{ij}/\sigma_i$, $\tilde{b}_i=b_i/\sigma_i$.
Every noise term then has unit variance, and the target no longer depends on the noise level,
\begin{equation}
   ||\tilde{A}x_\gamma - \tilde{b}||^2 \approx N_m .
\end{equation}
This also means that measurements with large variances are down-weighted in the inversions.
In the general case, the full noise covariance matrix should be considered, while our assumption of uncorrelated noise only considers the diagonal terms.
Estimating the off-diagonal terms from our limited measurements would be very uncertain, however.
Because the residual increases monotonically with $\gamma$, the matching value can be found with a simple bisection search in $\log\gamma$, which keeps the number of inversions required small.
Noise weighting rescales the forward operator relative to the regularization term, which changes the numerical value of the regularization parameter that corresponds to a given amount of smoothing.
We therefore quote regularization parameters for weighted inversions in normalized form, $\gamma = \gamma_{\rm w}/\overline{\sigma^{-1}}^{\,2}$, where $\gamma_{\rm w}$ is the parameter used in the weighted system and $\overline{\sigma^{-1}}$ is the mean of $1/\sigma_i$ over all measurements.
This makes them comparable to $\gamma$ in unweighted systems.
The correspondence is exact for uniform uncertainties and approximate otherwise.

Figure~\ref{fig:double_cell__inversion_of_forward_model_with_noise__different_regularization_operators} shows results from such noise-weighted inversions, this time using a forward model for the double-cell flow, also with 5\% noise added.
We consider here several choices of the regularization operator, $\Gamma$, and in each case select $\gamma$ with the discrepancy principle.
The operators range from the simplest, the identity matrix, $I$, for which the regularization term becomes the flow amplitude, to radial derivatives, $\frac{\partial}{\partial r}$, and the Laplacian, $\Delta$.
The derivative operators are expressed in matrix form using 4th-order finite differences with the Python package \emph{findiff} \citep{findiff}.
We also explored varying the effective regularization strength smoothly through the convection zone by multiplying $\Delta$ by a scaling function $s_\beta(r) = \beta^{(r-R_T)/(2(R_\sun-R_T))}$, where $\beta>1$ corresponds to stronger regularization near the solar surface than near the tachocline at $R_T=0.71R_\sun$.
As Figure~\ref{fig:double_cell__inversion_of_forward_model_with_noise__different_regularization_operators} illustrates, the choice of regularization operator can have a significant impact on the solution.
Surprisingly, radial derivatives and the Laplacian yielded very similar inversions, while the Laplacian with a depth-varying scaling function greatly improved correspondence with the true flow model, and in particular showed improved recovery of the deep flow reversal.
For the rest of this paper, we therefore consider only $\Gamma=s_{10}\Delta$.
In future work, one may choose scaling functions such that errors become evenly distributed over depths (corresponding travel distances).
 
\subsection{Multi-Component and Constrained Inversions} \label{sec:inversion_procedure:constrained_inversions}

\begin{figure}[t]
   \centering
   \vspace*{5mm}
   \includegraphics[width=0.85\linewidth]{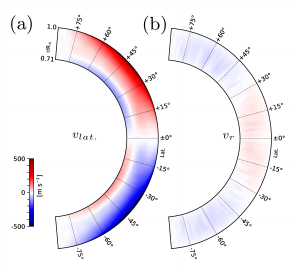}
   \caption{Inverted single-cell flow under mass conservation using the full forward model (latitudinal and radial contributions) with added Gaussian noise as synthetic measurements.
   Using the same color scale, panels (a) and (b) show latitudinal and radial flow velocities, respectively.
   Regularization operator for this case is the depth-scaled Laplacian $s_{10}\Delta$ and a regularization parameter $\gamma=5.6\times10^7$ was chosen using the L-curve criterion.
   \label{fig:single_cell__inversion_of_forward_model_with_noise__with_mass_conservation}}
\end{figure}

\begin{figure}[t]
   \centering
   \vspace*{5mm}
   \includegraphics[width=\linewidth]{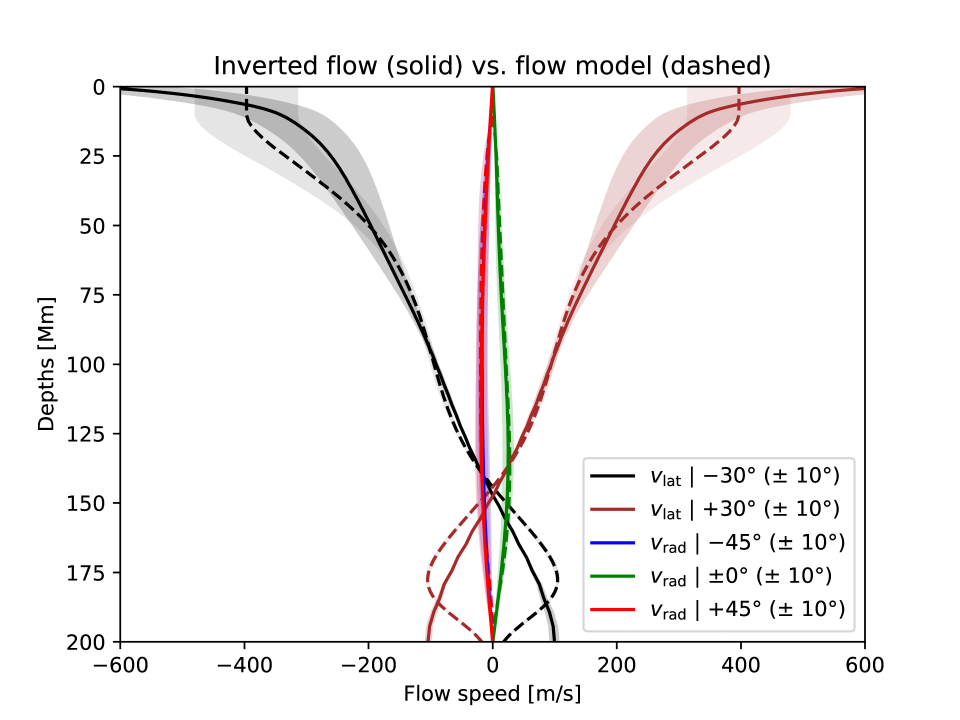}
   \caption{Comparing flow model with inversion of the single-cell forward model under mass conservation from Figure~\ref{fig:single_cell__inversion_of_forward_model_with_noise__with_mass_conservation}.
   Shown are profiles of the flow averaged over latitude bands, where solid curves denote the inversion result, and dashed curves the flow model.
   Shading shows the variation (one standard deviation) within those latitude bands.
   \label{fig:single_cell__inversion_of_forward_model_with_noise__with_mass_conservation__profiles}}
\end{figure}

Inversions can be extended straightforwardly to multiple components--such as different flow directions or sound-speed perturbations--and to multiple measurement types, including different geometries and frequencies.
In practice, this simply amounts to stacking the corresponding sensitivity kernels and measurement vectors within a single linear system.
In such cases, it is often advantageous to impose additional constraints on the solution.
A particularly useful class of constraints are equality constraints, which enforce that the solution satisfies prescribed relations, for example mass conservation.
These constraints can be incorporated using Lagrange multipliers, defining the augmented functional
${\mathcal {L}}(x,\lambda)=f(x)+\lambda^T g(x)$ where $f(x)=1/2||Ax-b||^2 + 1/2\gamma||\Gamma x||^2$ is the regularized least-squares problem to be minimized and $g(x)=Cx - c$ is the constraint.
Requiring stationarity of ${\mathcal {L}}$ with respect to $x$ and $\lambda$ yields:
\begin{eqnarray}
0 &=& A^T A x + \gamma \Gamma^T \Gamma x + C^T \lambda - A^T b\\
0 &=& C x - c
\end{eqnarray}
which can be written in block matrix form as
\begin{equation}
\left[ {\begin{array}{cc}
   A^T A + \gamma \Gamma^T \Gamma & C^T \\
   C & 0 \\
 \end{array} } \right]
 \left[ {\begin{array}{c}
    x \\
    \lambda \\
  \end{array} } \right] =
  \left[ {\begin{array}{c}
     A^T b \\
     c \\
   \end{array} } \right].
\end{equation}

To improve the numerical scaling of the augmented system, we scale the constraint operator $C$ such that its norm is comparable to that of the regularized normal-equations block $M=A^TA + \gamma \Gamma^T \Gamma$.
Specifically, we apply the rescaling $C \to sC$, $c \to sc$ with $s=||M||/||C||$ where the norms $||M||$ and $||C||$ denote Frobenius norms.
This balancing reduces numerical disparities between the matrix blocks and improves the stability and accuracy of the linear solve, while leaving the constrained solution itself unchanged (the Lagrange multipliers rescale accordingly).
Because the augmented system constitutes a saddle-point problem rather than a pure least-squares system, we solve it directly using standard linear-system techniques rather than the iterative LSQR procedure employed elsewhere in this work.
As a consequence, the constrained inversions are computationally more demanding than the unconstrained LSQR inversions, and have much higher memory requirements.
The constrained inversions shown throughout are therefore computed at one half the nominal resolution, that is, using 123 points in latitude and 87 points in depth.

Figure~\ref{fig:single_cell__inversion_of_forward_model_with_noise__with_mass_conservation} shows an inversion in which mass conservation is enforced in the simultaneous inference of latitudinal and radial flow components.
Mass conservation requires that both the divergence of the mass flow vanishes ($\rho\vec{v}=0$), and that there are no flows in or out of the domain.
The former we implement as a strict equality constraint, where the corresponding differential operator is again discretized on our spatial grid using fourth-order finite differences.
On the other hand, we implement the boundary conditions as soft penalty terms since strict enforcement can lead to poorly conditioned systems. 
In this sense, they act effectively as an additional regularization term, trading satisfaction of the boundary condition for improved stability of the inversion.

A comparison between the inverted flow and the actual flow model is shown in Figure~\ref{fig:single_cell__inversion_of_forward_model_with_noise__with_mass_conservation__profiles} which plots latitudinally-averaged flow profiles for both.
Considering the radial flow is more than an order of magnitude smaller than the latitudinal flow, it is well recovered in the inversion.
For instance, for the latitude band at $0\degr$ (averaged over $\pm 10 \degr)$, the peak inverted radial flow is 21\,m/s compared to 27\,m/s for the model.
The latitudinal flow is well recovered in the interior in terms of flow amplitude, in particular between depths of 50-150\,Mm, and the location of the flow reversal within $\lesssim 3$\,Mm.
However, larger discrepancies are found near the upper and lower boundary of the domain.
The near-surface of the inversion shows a strong gradient which does not match the flow model.
Since in contrast inversions of the noise-free forward model without an explicit regularization term (Figure~\ref{fig:differentcutoffparameters}) do not show this behavior and match the flow model extremely well even at the surface, we conclude that this is a bias introduced by the regularization.

\subsection{Averaging Kernels} \label{sec:results:averaging_kernels}

\begin{figure*}[p]
   \centering
   \includegraphics[width=\linewidth]{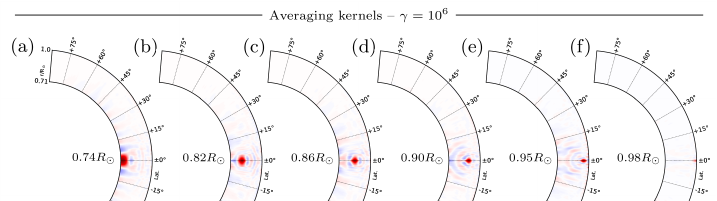}
   \includegraphics[width=\linewidth]{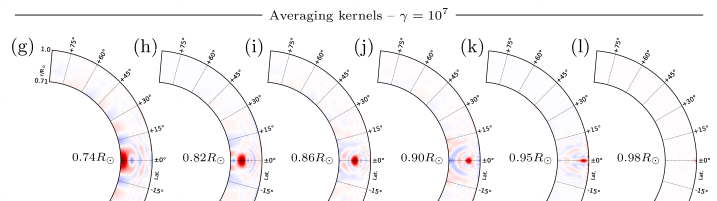}
   \includegraphics[width=\linewidth]{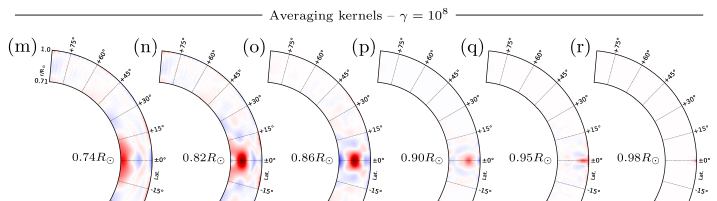}
   \caption{Averaging kernels for three values of the regularization parameter $\gamma$ (from top to bottom), for inversions with the depth-scaled Laplacian $s_{10}\Delta$ as the regularization operator.
   The plots show the contribution to the inferred latitudinal flow resulting from a true flow located at zero latitude and depths 0.74, 0.82, 0.86, 0.90, 0.95 and $0.98R_\sun$ (from left to right). 
   Each plot's color range is scaled to their respecive $\pm$ absolute maximum value which does vary between cases.
   Only half of the domain is shown since the kernels are symmetric between northern and southern hemispheres.
   \label{fig:averaging_kernels}}
\end{figure*}

So-called averaging kernels indicate how an inferred flow at some location relates to the true flow field.
Some local inversion procedures such as OLA (Optimally Localized Averages) and SOLA (Subtractive Optimally Localized Averages) \citep[e.g.][]{PijpersThompson1994} explicitly construct averaging kernels in such a way that they are close to some localized target distribution.
Our inversion procedure does not directly involve such kernels, and instead solves the linear system globally in a least-squares sense using an iterative procedure.

However, one may still want to compute effective averaging kernels as they reveal the effective spatial resolution and cross-depth leakage of the inversion, depending on the chosen regularization.
Averaging kernels depend only on the forward operator and the regularization, not on the data.
We compute them for the unweighted system, so that they do not depend on a noise model.
Let $S$ denote the augmented matrix of the system in Eq.~\ref{eqn:augmented_system}, and $S^+$ its Moore--Penrose pseudoinverse.
The regularized solution is then
\begin{equation}
  x_\gamma = S^+ \begin{bmatrix} b \\ 0 \end{bmatrix} = R\,b ,
\label{eqn:pseudo_inverse}
\end{equation}
where $R$ is the block of $S^+$ that acts on the data, i.e. its first $N_m$ columns.
Replacing $b$ by the noise-free forward solution $Ax$, where $x$ is the true flow field, we get
\begin{equation}
  x_\gamma = \mathcal{K}\,x, \qquad \mathcal{K} = R\,A ,
\end{equation}
where $\mathcal{K}$ is the matrix of averaging kernels: its $j$-th row describes how the true flow is averaged to produce the estimate at grid point $j$.

We compute the pseudoinverse by singular value decomposition (SVD), using the \emph{pinv} routine of the Python package \emph{SciPy}.
For efficiency and numerical stability, we ignore contributions with very small singular values.
A relative threshold of $rtol=10^{-3}$ (smallest/largest singular value considered) gives solutions close to those computed with LSQR using the $10^{-5}$ tolerance we have been using (see section~\ref{sec:inversion_procedure}).
Computing the pseudoinverse is significantly more expensive than LSQR, both in compute time and especially in memory, and we therefore use it only for select cases to compute averaging kernels, and otherwise use the LSQR-based solver.

Averaging kernels do depend on the type and amount of regularization.
Figure~\ref{fig:averaging_kernels} shows examples for the inversion of latitudinal flows, for the depth-scaled Laplacian regularization operator $s_{10}\Delta$, and for three values of $\gamma$.
As one would expect, larger regularization parameters result in spatially-larger averaging areas.
The averaging kernels also strongly depend on depth.
Target locations close to the solar tachocline show spatially-extended averaging kernels, while those in shallower layers are compact.
The kernels show both positive and negative slide lobes with shell-like structure, which is not surprising given that the underlying sensitivity kernels consist of alternating banana-shell structures.
A consequence of these averaging kernels is that features of the true flow that are small compared to the width of the kernel are suppressed in the inversion.

\subsection{Uncertainty Estimation} \label{sec:inversion_procedure:uncertainty_estimation}

\begin{figure}[t]
   \centering
   \includegraphics[width=0.53\linewidth]{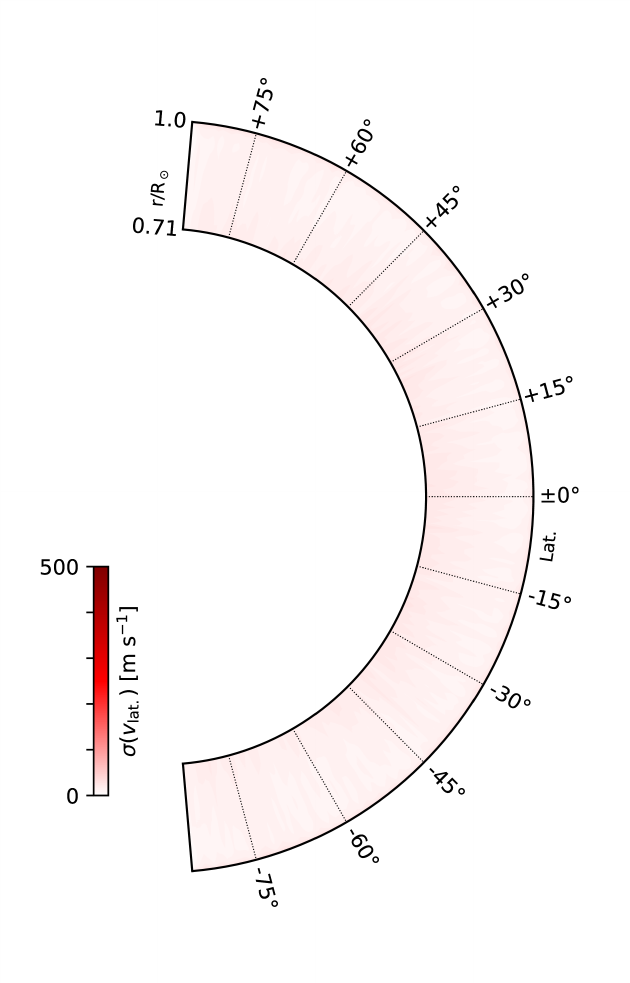}
   \caption{Standard deviation of the inferred flow due to propagated noise for the inversion of the double-cell forward model with noise from Figure~\ref{fig:double_cell__inversion_of_forward_model_with_noise__different_regularization_operators}(e), computed using LSQR from 50 Monte-Carlo samples, which limits the relative accuracy of the standard deviation to about 10\%.
   Color scale is the same as used for all flow models and inversion results throughput the paper, except that the standard deviation only has positive values.
   \label{fig:double_cell_uncertainty_estimation_forward}}
\end{figure}

The regularized solution $x_\gamma$ depends on the particular noise realization in the data.
It is therefore useful to quantify how much inversion results would vary for other realizations.
Propagating noise requires noise information, so we here consider the noise-weighted system, $\tilde{A}$ and $\tilde{b}$, in which the noise has unit variance in every component.
As shown above, for fixed $\gamma$ the solution is a linear function of the data, $x_\gamma = \tilde{R}\,\tilde{b}$, so a perturbation $\epsilon$ of the data changes the solution by $\tilde{R}\epsilon$, independent of the data themselves.
We therefore can estimate the uncertainty by Monte Carlo sampling: we draw $N_s=50$ independent vectors $\epsilon^{(k)}$ of unit-variance Gaussian noise, invert each with the same $\tilde{A}$, $\Gamma$, and $\gamma$ to obtain $\delta x^{(k)} = \tilde{R}\epsilon^{(k)}$, and take the standard deviation of the ensemble at each point as the uncertainty of $x_\gamma$.
The ensemble has zero mean by construction.
We evaluate this at the value of $\gamma$ selected for the actual data, so the uncertainty does not include the variation of the selected $\gamma$ between noise realizations, whether chosen via L-curve or discrepancy principle.
The spread measures only the propagation of noise.
It also does not include the bias introduced by the regularization itself, which is described by the averaging kernels above, so the true solution can lie outside the quoted uncertainty where the solution is strongly smoothed.
Figure~\ref{fig:double_cell_uncertainty_estimation_forward} shows the result of such calculations for the inversion of the double-cell forward model from Figure~\ref{fig:double_cell__inversion_of_forward_model_with_noise__different_regularization_operators}(e).
Latitudinal variations are minimal but there is a clear depth trend: at the base of the convection zone, the flow uncertainty is approximately 20~ms$^{-1}$, monotonically decreases to 10~ms$^{-1}$ at $0.95\,R_s\sun$, and then rises to the upper boundary to around 30~ms$^{-1}$.

\section{Inversion of synthetic measurements} \label{sec:synthetic}

While the forward model inversions in Section~\ref{sec:inversion_procedure} test the mathematical consistency of our inversion procedure, they do not independently validate the sensitivity kernels themselves, since the forward models were generated using the very same kernels used for the inversions.
In this section, we evaluate the complete methodology--including the simulation-based sensitivity kernels and the multi-frequency measurement procedure--by applying our framework to the travel-time measurements derived from simulated wavefields, as described in section~\ref{sec:models}.
All inversions in this section use the depth-scaled regularization operator $s_{10}\Delta$.

\subsection{Single-Cell Flow Recovery}

\begin{figure}[t]
   \centering
   \vspace*{5mm}
   \includegraphics[width=0.85\linewidth]{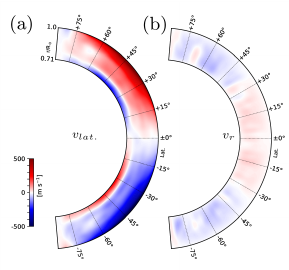}
   \caption{Inverted single-cell flow under mass conservation from measured travel-time shifts.
   Panels show (a) latitudinal and (b) radial flow velocities using the same color scale.
   Regularization operator for this case is the depth-scaled Laplacian $s_{10}\Delta$, with a regularization parameter $\gamma=3.2\times10^9$ selected by the L-curve criterion.
   \label{fig:single_cell__inversion_of_measurements__with_mass_conservation}}
\end{figure}

\begin{figure}[t]
   \centering
   \vspace*{5mm}
   \includegraphics[width=\linewidth]{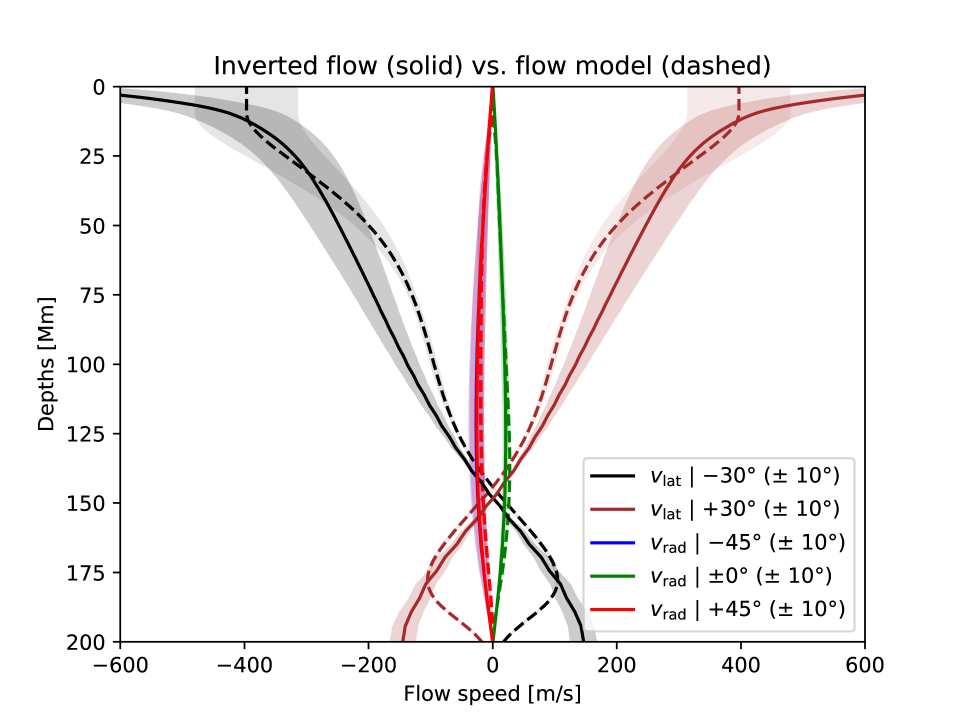}
   \caption{Comparing the single-cell flow model with the measurement inversion from Figure~\ref{fig:single_cell__inversion_of_measurements__with_mass_conservation}.
   Shown are profiles of the flow averaged over latitude bands where solid curves denote the inversion result, and dashed curves the flow model.
   Shading shows the variation (one standard deviation) within those latitude bands.
   \label{fig:single_cell__inversion_of_measurements__with_mass_conservation__profiles}}
\end{figure}

Figure~\ref{fig:single_cell__inversion_of_measurements__with_mass_conservation} presents results from the joint inversion of latitudinal and radial flows using the travel-time differences measured from the single-cell simulation, and using the 3\,mHz frequency bin.
This inversion uses mass conservation as a strict equality constraint and a zero-in/outflow boundary condition as a soft penalty term (see section~\ref{sec:inversion_procedure:constrained_inversions}).
A regularization parameter of $\gamma=3.2\times10^9$ was chosen using the L-curve criteria, since we do not have uncertainty estimates for this set.
Note that the value of $\gamma$ for this constrained single-cell inversion is not directly comparable to those of unconstrained inversions: the constrained problem solves for both latitudinal and radial flow, with a regularization term acting on both components, and imposes mass conservation through an augmented system with Lagrange multipliers.
Depth profiles of the flow model and inverted flow averaged over different latitude bands are compared in Figure~\ref{fig:single_cell__inversion_of_measurements__with_mass_conservation__profiles}, showing that results are in many aspects very similar to the inversion of the forward model.
The location of the flow reversal is well recovered, the crossover is $\sim4\,$Mm deeper in the inverted flow compared to the model.
Sign and amplitude of the radial flow match the model well.
For the latitude band at $0\degr$ (averaged over $\pm 10 \degr)$, the peak inverted radial flow is 21\,m/s vs. 27\,m/s for the model.
Larger discrepancies are found in the latitudinal flow, in particular at the top and bottom of the domain.
At least for the differences at the bottom, a likely contributor is the loss of resolution at depth, as illustrated by the averaging kernels in section~\ref{sec:results:averaging_kernels}.
But note that those kernels were computed for unconstrained inversions of a single field, so they qualitatively indicate how resolution degrades with depth rather than than giving the exact resolution of the constrained inversion shown here.

\subsection{Double-Cell Flow Recovery}

\begin{figure*}[t]
   \centering
   \includegraphics[width=\linewidth]{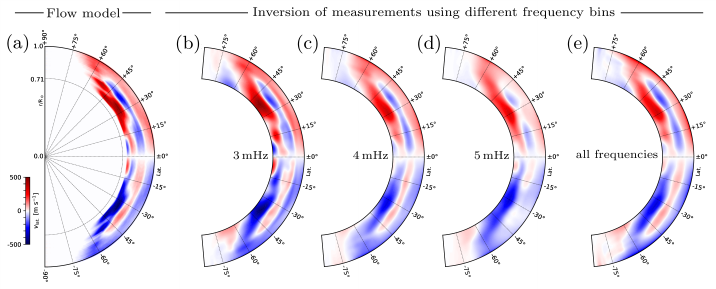}
   \caption{Inversion of travel-time differences measured from a numerical simulation of wave propagation in a solar model with a prescribed double-cell flow.
   The model is shown in panel (a), while panels (b)--(d) show inversion results using three different frequencies bins separately, and panel (e) shows results from a joint inversion using all three bins.
   Again, all panels use the same color scale.
   For the individual frequency inversions, regularization is chosen by the discrepancy principle with parameters $\gamma = 1.3\times10^7$, $8\times10^7$ and $7.3\times10^7$ for 3, 4 and 5\,mHz, respectively.
   The joint-frequency inversion incorporates an additional systematic noise term $\sigma_\mathrm{sys} = 0.25$\,s and a $\gamma = 1.6\times10^8$ (see text for details).
   \label{fig:double_cell__inversion_of_measurements_different_frequencies}}
\end{figure*}

\begin{figure}[t]
   \centering
   \vspace*{5mm}
   \includegraphics[width=\linewidth]{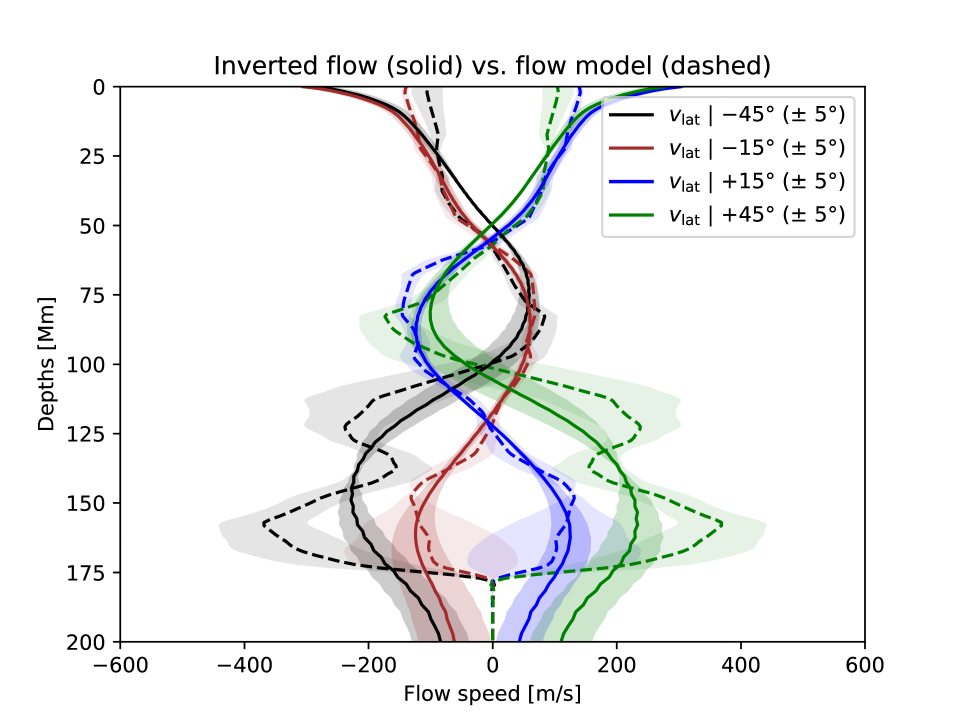}
   \caption{Comparing the double-cell flow model with the inversion of measurements incorporating all three frequency bins from Figure~\ref{fig:double_cell__inversion_of_measurements_different_frequencies}(e).
   Shown are profiles of the flow averaged over latitude bands, where solid curves denote the inversion result, and dashed curves the flow model.
   Shading shows the variation (one standard deviation) within those latitude bands.
   \label{fig:double_cell__inversion_of_measurements__profiles}}
\end{figure}

For the double-cell flow, we performed inversions for all three frequency bins, individually and jointly.
Figure~\ref{fig:double_cell__inversion_of_measurements_different_frequencies} shows results from these noise-weighted inversions.
The individual frequency inversions use the discrepancy principle to select the regularization that matches the noise level in the measurements, and all three produce sensible solutions with $\gamma = 1.3\times10^7$, $8\times10^7$ and $7.3\times10^7$ for 3, 4 and 5\,mHz, respectively.

The higher-frequency measurements are noisier, therefore the 4\,mHz and 5\,mHz data require stronger regularization and yield smoother inversions than the 3\,mHz data.
The L-curves of the individual inversions have only weakly defined corners and are used here merely as a consistency check.
The L-curve corner coincides with the discrepancy-principle value at 3\,mHz.
At 4\,mHz and 5\,mHz it lies at $\gamma=10^7$, about the same $\gamma$ as for 3\,mHz, whereas the discrepancy principle favors a roughly eight times larger $\gamma$ ($\approx 2.8$ in $\gamma^{1/2}$).
This is expected: with $\gamma$ normalized by the mean weight, the L-curve is largely insensitive to the overall noise amplitude, while the discrepancy principle is not.

In contrast, for the joint inversion, where we use the kernels and measurements of all three frequencies to invert for a common latitudinal flow, the discrepancy principle fails.
The weighted misfit reaches the noise expectation only for very small regularization ($\gamma \lesssim 10^4$), where the solution is unphysical and clearly overfitted.
The L-curve, on the other hand, still shows a corner at a regularization $\gamma = 1.8\times10^8$, which corresponds to a reasonable solution.
The failure is therefore not in the regularization itself: at the corner the weighted misfit still exceeds the expected value by a factor of $2.5$, so the data cannot be fit to the stated noise level at any sensible $\gamma$.
This points to the noise model, which includes only stochastic uncertainties estimated from the measurements.
The most likely missing contribution is uncertainty in the kernels.

We therefore add a constant systematic uncertainty $\sigma_\mathrm{sys}$ in quadrature to the uncertainty of every measurement, $\sigma_\mathrm{eff}^2 = \sigma^2 + \sigma_\mathrm{sys}^2$, as a crude, first-order account of the kernel uncertainty.
We choose $\sigma_\mathrm{sys} = 0.25$\,s such that the discrepancy limit is reached at the L-curve corner.
The agreement of the two criteria is thus by construction, but the corner itself barely moves when $\sigma_\mathrm{sys}$ is added ($1.8\times10^8 \to 1.6\times10^8$), which shows that the choice of $\gamma$ is insensitive to the noise model.
The misfit is not evenly distributed among the three sets: the weighted misfit divided by the number of measurements is $1.34$, $0.70$ and $0.94$ for 3, 4 and 5\,mHz.
A single constant $\sigma_\mathrm{sys}$ is therefore only an approximate remedy, and should be understood as an effective average value rather than a measured kernel error, which is likely correlated and frequency dependent.
The joint inversion shown in Figure~\ref{fig:double_cell__inversion_of_measurements_different_frequencies}(e) is then best viewed as a regularized compromise between the three data sets, not as a fit of each set to its noise level.

Figure~\ref{fig:double_cell__inversion_of_measurements__profiles} compares select depth profiles from this inversion at different latitudes with the flow model.
The depth of the flow reversal is recovered well, depending on latitude with 1-7\,Mm, both for the upper meridional cell and the second cell in the deep interior.
The profiles also show similar trends to those seen before.
There is a near-surface gradient that does not match the original flow model, which our regularized inversions generally produce.
Near the bottom of the domain, the flow model sharply reaches zero, which the inversion cannot match since regularization results in wide spatial averaging at such depths (again, see section~\ref{sec:results:averaging_kernels} for representative averaging kernels).

\section{Summary and Conclusions}\label{sec:summary}

We have presented simulation-based sensitivity kernels for travel-time measurements of meridional flow at several temporal frequencies (3, 4, and 5\,mHz), for both the latitudinal and the radial flow components, and tested them by inverting forward-modeled and measured travel-time shifts (from numerical simulations) for prescribed single-cell and double-cell flows.

The kernels reproduce the familiar banana-doughnut structure of Born-approximation kernels: for horizontal flow, the sensitivity is concentrated in a region around the ray path, with reduced sensitivity along the ray path itself and sidelobes further out.
At higher temporal frequencies, the primary region of sensitivity and its sidelobes are more finely spaced, which suggests potentially higher spatial resolving power.
However, this advantage is offset in practice by the smaller amplitude of higher-frequency waves, which makes real-world measurements noisier.
In our double-cell test, this is reflected in the stronger regularization required for the 4 and 5\,mHz data.
The radial-flow kernels, in contrast, show no comparable reduction along the ray path and may instead be strongest near it, closer to the ray-path picture.
The physical origin of this difference between horizontal and radial flow sensitivities remains to be investigated, but it may provide insight into how finite-wavelength effects modify the sensitivity to different flow components.

We performed two kinds of inversions.
For the single-cell flow, we use the 3\,mHz measurements to invert jointly for the full meridional flow, i.e., the latitudinal and radial components, using mass conservation as a constraint.
The inversion of the measurements recovers the sign and approximate amplitude of the radial flow, even though it is almost 20 times weaker than the latitudinal flow.
For the double-cell flow, we invert only for the latitudinal flow, and compare inversions of the three frequency bins, individually and jointly.
In both cases, the depth of the flow reversals in the latitudinal flow is recovered to within a $\sim4$\,Mm for the single cell and 1--7\,Mm for the double cell, including the deeper second cell.
The main discrepancies are a near-surface gradient not present in the flow models and the flow near the bottom of the domain, where the flow model drops sharply to zero but the averaging kernels are broad.

Combining frequencies in a joint inversion exposed an incompleteness in the noise model.
With stochastic uncertainties only, the discrepancy principle fails for the joint inversion: at the L-curve corner the weighted misfit is about 2.5 times the expected value.
We attribute this to kernel errors, which the stochastic uncertainties do not include, and to which a joint inversion with a single common model is more sensitive than the individual inversions.
Adding a constant systematic uncertainty of $\sigma_\mathrm{sys}=0.25$\,s brings the discrepancy limit to the L-curve corner, which itself barely moves.
This is a crude correction, however: it does not distribute the misfit evenly among the frequencies, which suggests that the systematic errors are frequency dependent.
Our joint-frequency inversion therefore represents a regularized compromise between the three data sets, with an effective $\sigma_\mathrm{sys}$ calibrated on the present data that should be reevaluated for other flows and for real observations.

Natural next steps are to consider frequency-dependent or signal-proportional systematic error terms, a correlated covariance matrix in place of the diagonal terms, and joint estimation of the regularization and systematic uncertainty,
with the hope is that this would allow multi-frequency measurements to be exploited for higher resolution.

\par\addvspace{2\baselineskip}
\section*{Acknowledgments}
This work was supported by NASA's Heliophysics Supporting Research program under grants 80NSSC19K0857 and 80NSSC24K0261, and NASA's SDO/HMI project.
Computational resources supporting this work were provided by the NASA High-End Computing (HEC) Program through the NASA Advanced Supercomputing (NAS) Division at Ames Research Center.

\bibliographystyle{aasjournal}
\bibliography{heliophysics_bibliography}

\end{document}